\documentclass[twoside]{article}
\usepackage{fleqn,espcrc2}

\title{Confining Configurations in QCD and Relation to Rigid
Strings}

\author{R.Parthasarathy 
\address{e-mail: sarathy@imsc.ernet.in} \\ The Institute of Mathematical
Sciences, \\  
        Chennai 600113, India.
} 
       
\begin{document}

\begin{abstract}
The gauge field configurations of QCD gauge fields in the
infrared regime are obtained by magnetic symmetry condition.
The effective dual action exhibits dual Meissner effect with
quarks included. A string representation of this action
corresponds to rigid string. 
\vspace{1pc}
\end{abstract}

\maketitle

\section{Introduction}

A description of confinement in QCD requires the presence of
magnetic monopoles\cite{thooft}. The resulting effective
theory can be interpretted as producing dual Meissner effect
\cite{nambu} responsible for confinement. Lattice QCD studies
indicate that monopoles cover the physical vacuum in the
confining phase\cite{kronfeld}. Confinement is in the infrared
regime where the QCD coupling is large so that one has to use
non-perturbative analysis. In the non-confining phase, in the
ultra-violet regime, we realize the important property,
'asymptotic freedom', by using the full $SU(3)$ gauge field
configurations $A^a_{\mu}(x)$. In the confining phase, we have
the monopole dominance and a partial (subset of)
$A^a_{\mu}(x)$ will be sufficient to realize confinement
through dual Meissner effect. This will be best described by
the 'magnetic symmetry' \cite{cho}, which reveals the
topological nature (of monopoles) of QCD. This symmetry
restricts the dynamical degrees of freedom while keeping the
gauge degrees of freedom.  

\section{Confining gauge field configurations}

The 'gauge covariant magnetic symmetry' condition for $SU(3)$
is
\begin{equation}
D^{ab}_{\mu}{\omega}^b\equiv {\partial}_{\mu}{\omega}^a+
gf^{acb}A^c_{\mu}{\omega}^b\ =\ 0,
\end{equation}
where ${\omega}^a(x)\in SU(3)$ and are chosen as
\begin{equation}
\sum_{a=1}^{8}{\omega}^a{\omega}^a=1 ; \sum_{b,c=1}^{8}
d^{abc}{\omega}^b{\omega}^c=\frac{1}{\sqrt{3}}{\omega}^a,
\end{equation}
$d^{abc}$ and $f^{abc}$ are the symmetric and anti-symmetric
tensors of $SU(3)$ respectively. These two conditions on the
octet vector ${\omega}^a$ are $SU(3)$ invariant. A general
solution to (1) is obtained as
\begin{equation}
A^a_{\mu}(x)=C_{\mu}(x){\omega}^a-\frac{4}{3g}f^{abc}
{\omega}^b({\partial}_{\mu}{\omega}^c),
\end{equation}
where $C_{\mu}(x)$ is an arbitrary vector field, independent
of ${\omega}^a$. The above gauge field configuration will be
taken here to describe QCD in the confining phase and will be
shown to produce dual Meissner effect and as well as the
confining rigid string action. By letting
${\omega}^a{\lambda}^a\ =\ \Omega$ (${\lambda}^a$'s are the
Gell-Mann matrices), with the conditions (2), the eigenvalues
of $\Omega$ are
$-\frac{1}{\sqrt{3}},-\frac{1}{\sqrt{3}},\frac{2}{\sqrt{3}}$.
There are two distinct eigenvalues and therefore
\cite{corrigan} the magnetic symmetry is governed by the
little group $U(2)$ of $\Omega$. The field strength associated
with (3) is found to be
\begin{equation}
F^a_{\mu\nu}=({\partial}_{\mu}C_{\nu}-{\partial}_{\nu}C_{\mu})
{\omega}^a-\frac{4}{3g}f^{abc}({\partial}_{\mu}{\omega}^b)
({\partial}_{\nu}{\omega}^c),
\end{equation}
and is "$SU(3)$ parallel" to ${\omega}^a$, that is, $f^{abc}
F^b_{\mu\nu}{\omega}^c\ =\ 0$ consistent with the identity
$[D_{\mu},D_{\nu}]^{ab}{\omega}^b=if^{abc}F^b_{\mu\nu}{\omega}
^c$, in view of (1). However, this does not imply that
$F^a_{\mu\nu}$ is along ${\omega}^a$.   

\section{Effective Dual Action}

Using the above field strength and the gauge field
configuration (3), the standard QCD action with the quarks
becomes
\begin{eqnarray} 
\lefteqn{S=-\frac{1}{4}\int\{f^2_{\mu\nu}-\frac{8}{3g}f_{\mu\nu}X
_{\mu\nu}+O(\frac{1}{g^2})\}d^4x}  \nonumber \\
\lefteqn{+\int\{\bar{\psi}i{\gamma}^{\mu}{\partial}_{\mu}\psi
-\frac{g}{2}\bar{\psi}{\gamma}^{\mu}{\lambda}^aC_{\mu}
{\omega}^a\psi\}d^4x}   \nonumber \\
\lefteqn{+\int\{\frac{2}{3}\bar{\psi}{\gamma}^{\mu}{\lambda}^a\psi f^{abc}
{\omega}^b({\partial}_{\mu}{\omega}^c)\}d^4x,}
\end{eqnarray}
where we have denoted
${\partial}_{\mu}C_{\nu}-{\partial}_{\nu}C_{\mu}$, the Abelian
part of $F^a_{\mu\nu}$ by $f_{\mu\nu}$ and $f^{abc}{\omega}^a
({\partial}_{\mu}{\omega}^b)({\partial}_{\nu}{\omega}^c)$, the
topological part of $F^a_{\mu\nu}$ by $X_{\mu\nu}$. It then
follows ${\partial}_{\mu}X_{\mu\nu}\neq 0$, ${\partial}_{\mu}
{\tilde{X}}_{\mu\nu}\neq 0$, where ${\tilde{X}}_{\mu\nu}=
\frac{1}{2}{\epsilon}_{\mu\nu\alpha\beta}X_{\alpha\beta}$.
Further,
\begin{equation}
-\frac{2}{3}\oint {\epsilon}_{\mu\nu\alpha\beta}f^{abc}
{\omega}^a({\partial}_{\alpha}{\omega}^b)({\partial}_{\beta}
{\omega}^c)dx^{\mu}\wedge dx^{\nu},
\end{equation}
is a topological invariant \cite{marciano} and therefore
$X_{\mu\nu}$ can be taken to describe monopoles. 

Introducing ${\cal{G}}_{\mu\nu}$ as dual to $f_{\mu\nu}$ in
(4), the variation with respect to the $C_{\mu}$-field yields
an equation for ${\cal{G}}_{\mu\nu}$, which is solved to
obtain
\begin{eqnarray}
\lefteqn{{\cal{G}}_{\mu\nu}={\epsilon}_{\mu\nu\lambda\sigma}{\partial}
_{\lambda}{\tilde{A}}_{\sigma}+\frac{4}{3}X_{\mu\nu} 
+\frac{g}{2{\pi}^2}\int d^4y \frac{1}{{\mid x-y\mid}^4}}
\nonumber \\
\lefteqn{\{
(x-y)_{\mu}J_{\nu}(y)-(x-y)_{\nu}J_{\mu}(y)\},} 
\end{eqnarray}
where ${\tilde{A}}_{\sigma}$ is dual to $C_{\mu}$ and the
current
$J_{\mu}=\frac{1}{2}\bar{\psi}{\gamma}^{\mu}{\lambda}^a\psi
{\omega}^a$. It also follows from the variational equation
that the current $J_{\mu}$ is conserved,
${\partial}_{\mu}J_{\mu}=0$. Using this, we obtain the dual
action
\begin{eqnarray}
\lefteqn{S_d=\int [-\frac{1}{4}{\tilde{f}}^2_{\mu\nu}-\frac{4}{3}
{\tilde{A}}_{\sigma}{\partial}_{\lambda}{\tilde{X}}_{\lambda
\sigma}+\frac{4}{9}{\tilde{X}}_{\mu\nu}{\tilde{X}}_{\mu\nu}}
\nonumber \\ 
\lefteqn{-\frac{g}{4{\pi}^2}{\tilde{A}}_{\mu}{\tilde{J}}_{\mu} 
+\frac{g}{4{\pi}^2}{\tilde{X}}_{\mu\nu}{\tilde{Y}}_{\mu\nu}
+\bar{\psi} i{\gamma}^{\mu}{\partial}_{\mu}\psi 
}   \nonumber \\ 
\lefteqn{+\frac{g^2}{8{\pi}^2}\int d^4y J_{\mu}(x)(x-y)^{-2}J_{\mu}
(y)} \nonumber \\  
\lefteqn{-\frac{4}{3}j^a_{\mu}f^{abc}{\omega}^a{\partial}_{\mu}
{\omega}^c]d^4x,} 
\end{eqnarray}  
where ${\tilde{f}}_{\mu\nu} =
{\partial}_{\mu}{\tilde{A}}_{\nu}-{\partial}_{\nu}{\tilde{A}}_
{\mu}$, ${\tilde{J}}_{\sigma}={\epsilon}_{\mu\nu\lambda\sigma}
{\partial}_{\lambda}Y_{\mu\nu}$, the dual current, where
$Y_{\mu\nu}=\int \frac{d^4y}{{\mid x-y\mid}^4}\{(x-y)_{\mu}
J_{\nu}(y)-(x-y)_{\nu}J_{\mu}(y)\}$ and $j^a_{\mu}=\bar{\psi}
{\gamma}^{\mu}{\lambda}^a\psi$. 
In here, we have the dual
Abelian field ${\tilde{A}}_{\sigma}$ coupled to the dual
current ${\tilde{J}}_{\sigma}$ and a Biot-Savart energy term
for the quarks. All the fields are massless and the dual
action (6) has dual $U(1)$ gauge invariance.

\section{Generation of mass for dual Abelian field}

In order to realize dual Meissner effect, the dual gauge field
must acquire mass. Instead of introducing a scalar field with
its own potential, we generate a mass term by quantum
fluctuations and using 'gauge mixing mechanism'
\cite{aurilia}. We first consider the quantum fluctuations of
${\tilde{A}}_{\mu}$ and ${\tilde{X}}_{\mu\nu}$ and integrate
over the fluctuations of ${\tilde{A}}_{\mu}$. We do not
encounter Gribov problem as these fields are Abelian. The
integration of the fluctuations of the dual Abelian field
produces the following additional terms,
\begin{equation}
\frac{4}{9}{\tilde{x}}^2_{\mu\nu}+a({\partial}_{\mu}{\tilde{x}
}_{\mu\nu})^2-\frac{4}{3}{\tilde{A}}_{\sigma}({\partial}
_{\lambda}{\tilde{x}}_{\lambda\sigma}),
\end{equation}
where ${\tilde{x}}_{\mu\nu}$ is the fluctuation of
${\tilde{X}}_{\mu\nu}$ and $'a'$ is a dimensionful parameter
coming from the integration of the fluctuations of
${\tilde{A}}_{\mu}$. We now have a massive anti-symmetric
tensor field ${\tilde{x}}_{\mu\nu}$ coupled to massless dual
field ${\tilde{A}}_{\sigma}$. We still have dual $U(1)$ gauge
invariance. Representing ${\partial}_{\lambda}{\tilde{x}}_{
\lambda\sigma}$ by ${\partial}_{\sigma}\xi$, the equation of
motion
${\partial}_{\sigma}\xi=\frac{2}{3a}{\tilde{A}}_{\sigma}$, is
used to obtain the action
\begin{eqnarray}
\lefteqn{S_d=\int [ -\frac{1}{4}{\tilde{f}}^2_{\mu\nu}-\frac{4}{3}
{\tilde{A}}_{\sigma}{\partial}_{\lambda}{\tilde{X}}_{
\lambda\sigma}+\frac{4}{9}{\tilde{X}}^2_{\mu\nu}-\frac{g^2}
{4{\pi}^2}{\tilde{A}}_{\sigma}{\tilde{J}}_{\sigma}} 
\nonumber \\
\lefteqn{+\frac{g}{4{\pi}^2}{\tilde{X}}_{\mu\nu}{\tilde{Y}}_{\mu\nu}
-\frac{4}{9a}{\tilde{A}}^2_{\sigma}\ +\ Dirac terms\ ]d^4x,}  
\end{eqnarray}
thereby generating mass term for the dual Abelian field.

\section{Dual Meissner effect}

The equation of motion for ${\tilde{A}}_{\sigma}$ from (8) is
\begin{equation}
{\partial}_{\mu}{\tilde{f}}_{\mu\nu}=\frac{4}{3}{\partial}_{
\mu}{\tilde{X}}_{\mu\nu}+\frac{g^2}{4{\pi}^2}{\tilde{J}}_{\nu}
+m^2{\tilde{A}}_{\nu},
\end{equation}
where $m^2=\frac{8}{9a}$. It follows from the above expression
that ${\partial}_{\mu}{\tilde{A}}_{\mu}=0$. The above
expression is used to eliminate ${\tilde{A}}_{\mu}$ from (8)
to finally obtain an effective action in the infrared regime
as
\begin{eqnarray}
\lefteqn{S_{eff.}=\int[-\frac{8}{9}{\partial}_{\lambda}{\tilde{X}}
_{\lambda\nu}({\partial}^2-m^2)^{-1}{\partial}_{\rho}
{\tilde{X}}_{\rho\nu}+\frac{4}{9}{\tilde{X}}^2_{\mu\nu}
} \nonumber \\  
\lefteqn{+\frac{g^4}{32{\pi}^4}{\tilde{J}}_{\nu}({\partial}^2-m^2)
^{-1}{\tilde{J}}_{\nu}-\frac{g^2}{3{\pi}^2}{\tilde{J}}_{\nu}
({\partial}^2-m^2)^{-1}{\partial}_{\lambda}{\tilde{X}}_
{\lambda\nu}}  \nonumber \\  
\lefteqn{+\frac{g}{4{\pi}^2}{\tilde{X}}_{\mu\nu}{\tilde{Y}}_{\mu\nu}
+\bar{\psi}i{\gamma}^{\mu}{\partial}_{\mu}\psi 
} \nonumber \\ 
\lefteqn{+\frac{g^2}
{8{\pi}^2}J_{\mu}(x)
\int d^4y(x-y)^{-2}J_{\mu}(y)} \nonumber \\  
\lefteqn{-\frac{4}{3}j^{a\mu} f^{abc}
{\omega}^a ({\partial}_{\mu}{\omega}^c)]d^4x.}  
\end{eqnarray}
 
The first term is as in the London theory of Meissner effect
which confines the monopole configurations and the second term
corresponds to mass for these objects. The third term confines
quarks through ${\tilde{J}}_{\mu}$ by Dual Meissner mechanism.
In addition we have a coupling of quark current and monopole
configuration. The dual field strength ${\tilde{X}}_{\mu\nu}$,
which contains the monopoles, can be interpretted as Abrikosov
flux lines responsible for the confinement of quarks and they
themselves are confined. 

The structure of the quark current $J_{\mu}$ or its dual, has
${\lambda}^a{\omega}^a=\Omega$. In a representation in which
$\Omega$ is diagonal, this endows magnetic charges to the
coloured quarks. There are two magnetic charges corresponding
to the two distinct eigenvalues of $\Omega$. This allows the
construction of mesons and baryons with zero magnetic charge
and the magnetic charge neutrality of the physical hadrons
ensures the colour neutrality. 
 
\section{A string representation}  

To obtain a string representation of the above infrard
effective dual action, we consider the situation without
quarks. Quarks can be included directly to the resulting
string action, by coupling them to the worldsheet and this
just renormalizes the string couplings without introducing new
terms \cite{parthasarathy}. The monopole configuration
$X_{\mu\nu}$ is represented now by \cite{wentzel}
\begin{equation}
X_{\mu\nu}=\int d^2\xi\ {\delta}^4(x-y)\ [y_{\mu},y_{\nu}],
\end{equation}
where $y_{\mu}({\xi}^1,{\xi}^2)$ represents a point on the
worldsheet swept by the string of the monopole,
$[y_{\mu},y_{\nu}]={\epsilon}^{ab}{\partial}_ay_{\mu}\
{\partial}_by_{\nu}$, with
${\partial}_a=\frac{\partial}{\partial {\xi}^a}$ and it can be
seen that $[x_{\mu},x_{\nu}]^2=2g$ where $g$ is the
determinant of the induced metric (first fundamental form) on
the worldsheet, namely,
$g_{ab}={\partial}_ax^{\mu}{\partial}_bx^{\mu}$. Then using
the above representation for $X_{\mu\nu}$, we obtain the
string action
\begin{eqnarray}
\lefteqn{S_{string}\ =\ \frac{m^2}{9\pi}K_0(\frac{m}{2\Lambda}
)\int \sqrt{g}d^2\xi} \nonumber \\
\lefteqn{-\frac{{\Lambda}^2}{18\pi m^2}\int \sqrt{g}
g^{ab}{\partial}_at_{\mu\nu}\ {\partial}_bt_{\mu\nu}\ d^2\xi}
\nonumber \\
\lefteqn{+\frac{{\Lambda}^2}{m^2}\int \sqrt{g}R\ d^2\xi},
\end{eqnarray}
where $\Lambda$ is an ultra-violet cut-off, $K_0$ is the
modified Bessel function and $t_{\mu\nu}\ =\
\frac{1}{\sqrt{g}}[x_{\mu},x_{\nu}]$. The first term is the
Nambu-Goto area term in the string action, the second term is
the Polyakov-Kleinert extrinsic curvature (second fundamental
form) action \cite{polyakov}  
and the third one is the Euler characteristic of
the string world sheet. Thus the string representation of the
effective dual action derived here corresponds to rigid
string.


\begin{thebibliography}{10}
\bibitem{thooft} G.'tHooft, 
                 Nucl.Phys. B190 (1981) 455;
                 S.Mandelstam,
                 Phys.Rev. D19 (1978) 2391.
\bibitem{nambu}  Y.Nambu,
                 Phys.Rep. C23 (1975) 250.  
\bibitem{kronfeld} A.Kronfeld, G.Schierholz and U.J.Wiese,
                   Nucl.Phys. B293 (1987) 461.  
\bibitem{cho} Y.M.Cho,
              Phys.Rev. D21 (1980) 1080; D23 (1981) 2413.
\bibitem{corrigan} E.Corrigan and D.Olive,
                   Nucl.Phys. B110 (1976) 237.
\bibitem{marciano} W.J.Marciano and H.Pagels,
                   Phys.Rev. D12 (1975) 1093.
\bibitem{aurilia} A.Aurilia and Y.Takahashi,
                  Prog.Theor.Phys. 66 (1981) 693.
\bibitem{parthasarathy} R.Parthasarathy and K.S.Viswanathan,
                        Lett.Math.Phys. 48 (1999) 243.
\bibitem{wentzel} G.Wentzel,
                  Suppl.Prog.Theor.Phys. Nos.37,38 (1966) 163.
\bibitem{polyakov} A.M.Polyakov,
                   Nucl.Phys. B268 (1986) 406;  
                   H.Kleinert,
                   Phys.Lett. B174 (1986) 335.  
\end{thebibliography}
\end{document}